\documentclass{ws-ijmpa}

\begin{document}

\markboth{Authors' Names}
{Instructions for Typing Manuscripts (Paper's Title)}

%
\catchline{}{}{}{}{}
%

\title{Exclusive production of $\rho^0 \rho^0$ pairs \\
in ultrarelativistic heavy ion collisions
}

\author{MARIOLA KLUSEK-GAWENDA
}

\address{
Institute of Nuclear Physics PAN, ul. Radzikowskiego 152\\
PL-31-342 Cracow, Poland \\
Mariola.Klusek@ifj.edu.pl}

\author{ANTONI SZCZUREK
}

\address{
Institute of Nuclear Physics PAN, ul. Radzikowskiego 152\\
PL-31-342 Cracow, Poland, \\
University of Rzesz\'ow, ul. Rejtana 16
PL-35-959 Rzesz\'ow, Poland.
\\
Antoni.Szczurek@ifj.edu.pl}

\maketitle

\begin{history}
\received{Day Month Year}
\revised{Day Month Year}
\end{history}

\begin{abstract}
We disuss exclusive electromagnetic production 
of two neutral $\rho^0$~mesons and show the predictions for 
the $AA \to AA \rho^0 \rho^0$ reactions 
for gold-gold collisions at the energy of $\sqrt{s}$~=~200 GeV (RHIC) and 
for lead-lead collisions at the energy of $\sqrt{s}$~=~5.5~TeV (LHC).
The elementary cross section is calculated with the help of 
the vector-dominance-model (VDM)-Regge approach 
which usually very well describes the experimental data at large 
$\gamma \gamma$ energy. 
The low-energy $\gamma \gamma \to \rho^0 \rho^0$ cross section is parametrized.
The cross section for nuclear process is calculated by means of 
the equivalent photon approximation (EPA).
We compare the results with realistic charge density with the results
for monopole form factor.

\end{abstract}

\ccode{PACS numbers: 13.40.Gp, 13.60.Le, 14.40.-n, 24.10.-i}

\section{Introduction}
%
\begin{figure}[h]
\centerline{\psfig{file=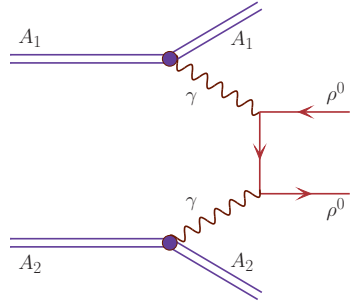,width=4.7cm}}
\vspace*{8pt}
\caption{The Born diagram for the exclusive $\rho^0$ pair production. 
\label{AAtoAArhorho}}
\end{figure}
Fig.~\ref{AAtoAArhorho} shows the basic mechanism of the exclusive
electromagnetic meson pair production. Two neutral $\rho$ mesons are
produced in coherent photon-photon processes in ultrarelativistic
heavy-ion collisions. So far only $AA \to AA \rho^0$ was measured
\cite{STAR_rho} and estimeted in the literature. We consider collisions
at RHIC and at LHC, where RHIC collides gold ions at the energy of 200 GeV
and LHC will collide soon lead nuclei at the energy 
of 5.5 TeV per nucleon. 

\section{Formalism}

EPA is very often used for calculating cross sections for 
electromagnetic interactions. Due to the coherent 
action of all the protons in the nucleus, the ions are surrounded by 
a strong electromagnetic field. This field can be viewed as a cloud 
of virtual photons which can collide with each other. 
This approach allows to consider production of mesons in peripheral 
ultrarelativistic nuclear collisions. ``Peripheral'' means that 
the distance between the nuclei is in practice bigger than
the sum of the radii of the two nuclei ($b>R_1+R_2\simeq 14$ fm).
The total cross section in EPA takes the form of the convolution of 
the elementary cross section ($\gamma \gamma \to \rho^0 \rho^0$) and 
the equivalent photon fluxes:
\begin{equation}
  \sigma\left(AA \rightarrow \rho^0 \rho^0  AA;s_{AA}\right) 
 = \int {\hat \sigma} \left( \gamma\gamma\rightarrow \rho^0 \rho^0 ;x_1 x_2 s_{AA} \right) {\rm d}n_{\gamma\gamma}\left(x_1,x_2,{\bf b}\right).
\label{EPA_1}
\end{equation}
We have introduced a new kinematical variable: $x=\frac{\omega}{E_A}$, 
where $\omega$ is the energy of the photon and the denominator is
the energy of the nucleus. After several transformation, we obtain 
the final form for the nuclear cross section in EPA:
\begin{eqnarray}
&  \sigma  & \left( AA \rightarrow \rho^0 \rho^0 AA;s_{AA} \right) = 
\int {\hat \sigma} \left(\gamma\gamma \rightarrow \rho^0 \rho^0;W_{\gamma \gamma} \right) \, \theta \left(|{\bf b}_1-{\bf b}_2|-2R_A \right) \nonumber \\
 & \times & N \left( \omega_1, {\bf b}_1 \right ) N \left( \omega_2, {\bf b}_1 \right ) 2 \pi b_m \, {\rm d} b_m \, {\rm d} \overline{b}_x \, {\rm d} \overline{b}_y \frac{W_{\gamma \gamma}}{2} {\rm d}W_{\gamma \gamma} {\rm d} Y  ,
 \label{EPA_end}
\end{eqnarray}
Here we put the $\theta$ function which assures only peripheral
collisions. 
In addition we define: $Y=\frac{1}{2} \left( y_{\rho^0} + y_{\rho^0} \right)$, $\overline{b}_{x/y} = \left( b_{1x/y} +b_{2x/y} \right)$ and $b_m=\overrightarrow{b_1} - \overrightarrow{b_2}$. 
The details of the derivation of Eq.~(\ref{EPA_end}) can be found in our
recent paper \cite{our_muon} on the muon pair production in 
the same approach.

The elementary cross section is divided into two components
\cite{our_rho}. 
The low-energy part is parametrized and the parameter are fitted to 
the $e^+ e^-$ data \cite{our_rho} while the high-energy part is obtained
with the help of the vector-dominance Regge type model with 
the parameters \cite{ASz_VDM} which are used to descibed other 
hadronic processes . 

The equivalent photon spectra (see Eq.~(\ref{EPA_end}) depend on 
the charge form factor of nuclei. The realistic form factor is 
the Fourier transform of the realistic charge distribution:
\begin{equation}
F(q) = \int \frac{4 \pi}{q} \rho \left( r \right) sin \left( qr \right) r dr.
\end{equation}
In the literature often a monopole form factor is used:
\begin{equation}
 F(q^2) = \frac{\varLambda^2}{\varLambda^2 + q^2}.
\end{equation}
The two form factors coincide only in a very limited range of $q$. 
With a larger value of the momentum transfer the difference between 
them becomes larger and larger. Inclusion of the realistic form factor
becomes particularly important at large meson pair rapidities
\cite{our_rho}. 
%
\section{Results and Conclusions}
%
\begin{figure}[h]
\center{\psfig{file=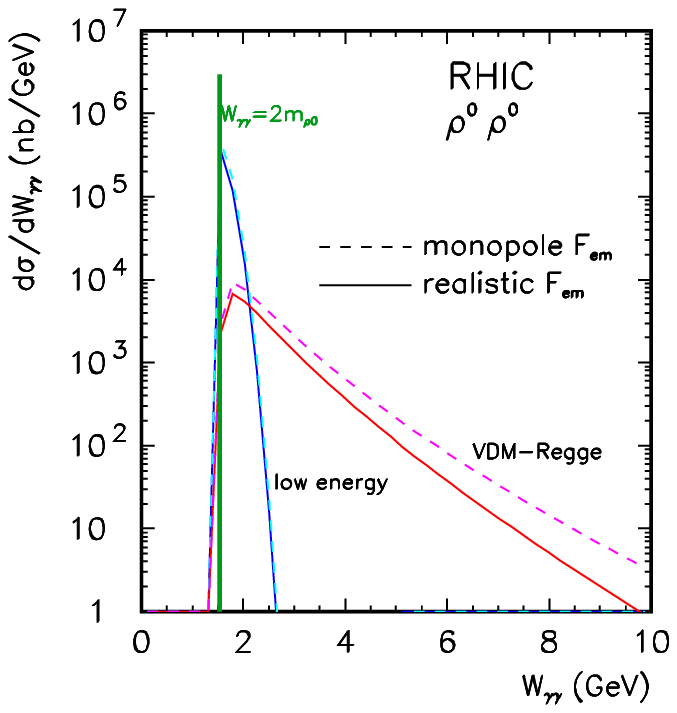,width=4.7cm}}
{\psfig{file=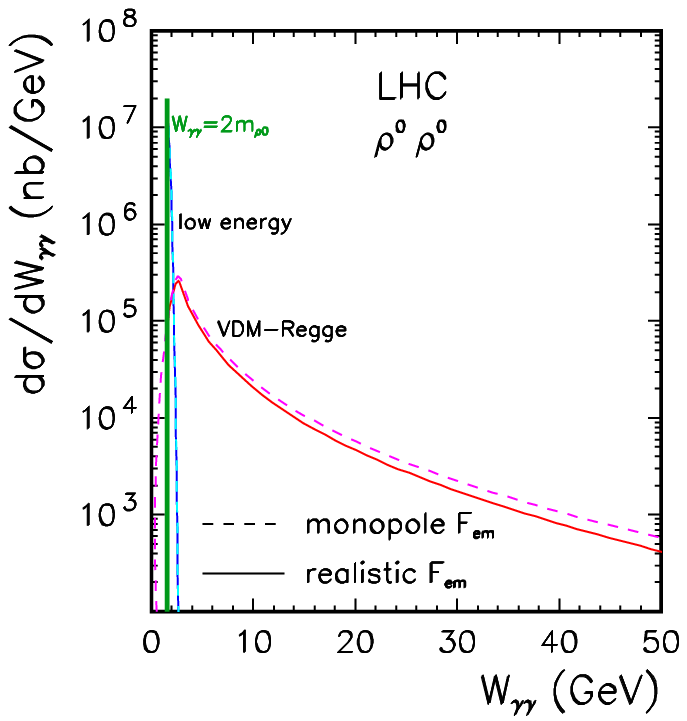,width=4.7cm}}
\vspace*{8pt}
\caption{The cross section for $Au-Au$ (left panel) and for $Pb-Pb$ scattering as a function of photon-photon center of mass energy $W_{\gamma \gamma}=M_{\rho^0 \rho^0}$. 
\label{ds_dw}}
\end{figure}
Fig. \ref{ds_dw} shows the distribution of the cross section for 
the nucleus-nucleus scattering in the photon-photon center of mass
energy $W_{\gamma \gamma}$ (i.e. the pair invariant mass) for both the
low-energy component and high-energy VDM-Regge component.
This distribution should be relatively easy to measure. 
The distributions in the impact parameter or in the rapidity of the pair
(not shown here) also show that the low-energy component is 
the dominant one. 
The cross section obtained with the help of the monopole form factor 
is larger than that obtained from the realistic charge form factor.
In addition we have observed that for smaller center-of-mass energy
the difference between the realistic and monopole form factor is bigger.
\\
\\
This work was partially supported by the Polish grant 
N N202 078735.
%


\begin{thebibliography}{0}    

\bibitem{STAR_rho}
The STAR Collab. (B.I. Abelev, {\it et al}.), 
{\it Phys. Rev. C\/} {\bf 77}, 034910 (2008).

\bibitem{our_muon} M. Klusek-Gawenda, A. Szczurek,
{\it Phys. Rev. C\/} {\bf 82}, 014904 (2010).

\bibitem{our_rho} M. Klusek, W. Schafer, A. Szczurek,
{\it Phys. Lett. B\/} {\bf 674}, 92-97 (2009).

\bibitem{ASz_VDM}
A. Szczurek, N.N. Nikolaev, J. Speth,
{\it Phys. Rev. C\/} {\bf 66}, 055206 (2002).

\end{thebibliography}
\end{document}